\documentclass[preprintnumbers,amsmath,amssymb]{
revtex4}
\usepackage{graphicx}
\usepackage{bm}

\newcommand{\w}{{\rm w}}
\newcommand{\nos}{\eta^2(0)}

\newcommand{\e}{\varepsilon}
\newcommand{\s}{\sigma}
\newcommand{\n}{\eta}
\newcommand{\om}{\omega}
\newcommand{\op}{\omega'}
\newcommand{\vf}{\vec{f}_k}
\newcommand{\vp}{\vec{\pi}}
\newcommand{\vx}{\vec{x}}
\newcommand{\vk}{\vec{k}}
\newcommand{\la}{\lambda}

\def \lta {\mathrel{\vcenter
     {\hbox{$<$}\nointerlineskip\hbox{$\sim$}}}}

\begin{document}

\preprint{UA-NPPS/07/2003}
\title{Non-equilibrium phenomena in the QCD phase transition}
\author{E.~N.~Saridakis}
\email{msaridak@phys.uoa.gr}
\author{N.~G.~Antoniou}
\email{nantonio@cc.uoa.gr}
\author{F.~K.~Diakonos}
\email{fdiakono@cc.uoa.gr}
\author{N.~Tetradis}
\email{ntetrad@cc.uoa.gr}
\affiliation{Physics Department, University of Athens, 15771 Athens, Greece}

\begin{abstract}
Within the context of the
linear $\s$-model for two flavours,
we investigate non-equilibrium phenomena that may 
occur during the QCD chiral phase transition in heavy-ion 
collisions. 
We assume that the 
chiral symmetry breaking is followed by a rapid
quench so that the system falls out of thermal equilibrium.
We study the mechanism 
for the amplification of the pion field during the oscillations of the
$\s$-field
towards and around its new minimum. We show that the pion spectrum 
develops a characteristic pronounced peak at low momenta.
\end{abstract}

\maketitle

\section{The model}

Experiments at RHIC and LHC are expected to probe many questions
in strong interaction physics.
One major area of interest concerns the chiral phase
transition. For given baryon-number chemical potential $\mu$
there exists a critical
temperature $T_{cr}$ above which the system lies in the
chirally symmetric state. As the temperature decreases below
$T_{cr}$ the system moves into the chirally broken phase.
It is believed that, for two flavours and zero quark masses, 
there is a 1st-order phase transition line
on the $(T,\mu)$ surface at large $\mu$ \cite{RW1}.
This line ends at 
a tri-critical point beyond which the phase transitions
become 2nd order. The line of 2nd-order transitions ends on the
$\mu=0$ axis. In the case of non-zero quark masses,
the 1st-order line ends at a critical point, beyond which 
the 2nd-order transitions are replaced by analytical crossovers.

This phase diagram has been discussed within various frameworks.
Our interest lies in the study
of possible non-equilibrium phenomena that may occur during the
phase transition. In particular we would like to study the possibility
that the system falls out of thermal equilibrium through rapid 
expansion. This is a realistic possibility in the
framework of heavy-ion collisions.

The scenario we have in mind assumes an initial thermalization at
a sufficiently high temperature for the system to move into the
chirally symmetric phase. 
The subsequent fast expansion generates deviations from thermal
equilibrium. We model this process by a quench during which the
volume of the system increases instantaneously by a certain
factor, with the number densities of the various particles
decreasing by the same factor.

We consider only the two lightest flavours and neglect the
effects of the strange quark.
As an effective description of the chiral theory
we use the $\s$-model \cite{GL,RW}.
The Lagrangian density is 
\begin{equation}
\mathcal{L}=\frac{1}{2}(\partial_\mu\s\partial^\mu\s
+\partial_\mu\vp\partial^\mu\vp)-V(\s,\vp)
\label{lagr}
\end{equation}
with the potential
\begin{equation}
V(\s,\vp)=\frac{\la^2}{4}(\s^2+\vp^2-v^2)^2
+\frac{m_{\pi}^2}{2}\left(\s^2+\vp^2-2v\s+v^2\right).
\label{pot}
\end{equation}
The last term in the potential accounts for     
the explicit chiral symmetry breaking by the quark masses.
The scalar field $\s$ together with the pseudoscalar
field $\vp=(\pi^+,\pi_0,\pi^-)$ form a chiral field $\Phi=(\s,\vp)$.
When the symmetry is restored at high temperatures
in the absence of the second term in the potential,
the system lies in the symmetric state
$\langle\s\rangle=\langle\vec{\pi}\rangle=0$. 
However, in the presence of the explicit symmetry breaking term in 
(\ref{pot}), 
the expectation value of the $\s$-field 
 never vanishes, so that the chiral symmetry is never completely restored.
At zero temperature and chemical potential
the expectation values of the fields become
$\langle\s\rangle=v=f_\pi$ and $\langle\vp\rangle=0$, where $f_\pi$
is the pion decay constant. 
We fix the parameters of the Lagrangian using the phenomenological
values 
$m_\pi$$\approx139\,$ MeV, 
$m_\s=\sqrt{2\lambda^2 f^2_\pi+m_\pi^2}\approx600\,$ MeV,
$v\approx87.4\,$ MeV, 
which yield $\la^2\simeq 20$.

\section{Equations of Motion}

The equations of motion resulting
from (\ref{lagr}) are:
\begin{eqnarray}
\partial_\mu\partial^\mu \s
+\la^2(\s^2+\vp^2-v^2)\s+m_\pi^2\s&=&vm_\pi^2\nonumber\\
\partial_\mu\partial^\mu \vp+\la^2(\s^2+\vp^2-v^2)\vp+m_\pi^2\vp&=&0.
\label{eomsfull}
\end{eqnarray}
We neglect the fluctuations of $\s$, while we treat $\vp(\vx,t)$ as a 
quantum field:
\begin{equation}
\vp(\vx,t)=\int\frac{d^3k}{(2\pi)^3}\left(a_{ k,v}\vf(t)\,e^{-i\vk\vx}
+a^\dag_{k,v}\vf^\star(t)\,e^{i\vk\vx}\right).
\label{pion}
\end{equation}
The creation and annihilation operators $a^\dag_{k,v}$,$a_{k,v}$ are
defined in the interaction picture at the vacuum corresponding
to the minimum $v$ of the potential, and $\vf(t)$ are the mode
functions 
of the pion field.

We work within the framework of the Hartree approximation. This leads
to
the effective replacement \cite{lenaghan}
\begin{itemize}
\item 
$\vp^2(\vx,t)\to\langle\vp^2(\vx,t)\rangle
=\langle\vp^2(t)\rangle$ 
\item
$\vp^2(\vx,t)\vp(\vx,t)\to\frac{5}{3}\langle\vp^2(t)\rangle\,\vp(\vx,t)$,
\end{itemize}
with 
$\langle \pi^2_i(t)\rangle= \langle\vp^2(t)\rangle/3$ for i=1,2,3.
The second approximation results from the 
replacement of the term $\pi_i^2\pi_j$
by $3\langle\pi^2_i\rangle\pi_j$ for $j=i$, and by 
$\langle\pi_i^2\rangle\pi_j$ for $j\not= i$.
Substituting the above approximations into (\ref{eomsfull}) we get:
\begin{equation}
\ddot{\s}(t)+\la^2\left(\s^2(t)+\langle\vp^2(t)\rangle-v^2\right)\s(t)
+m_\pi^2\s(t)=vm_\pi^2
\label{eqs}
\end{equation}
\begin{equation}
\vec{\ddot{f_k}}(t)+\left[k^2-\la^2v^2+\la^2\s^2(t)
+\frac{5}{3}\la^2\langle\vp^2(t)\rangle+m_\pi^2\right]\,\vf(t)=0.
\label{eqp}
\end{equation}
In (\ref{eqs}),(\ref{eqp}) $\langle\vp^2(t)\rangle$ is given by
\begin{equation}
\langle\vp^2(t)\rangle=\int\frac{d^3k}{(2\pi)^3}\vf^\star(t)\vf(t).
\label{pi2t}
\end{equation}

The particle density per momentum mode, for each component $i$ 
of the pion field is \cite{KLS,B}:
\begin{equation}
n^{}_{ki}=\frac{\omega_k}{2} \left( 
\left|f^{}_{ki}(t)\right|^2
+\frac{\left|\dot{f}^{}_{ki}(t)\right|^2}{\omega_k^2}\right)
-\frac{1}{2},
\label{nki}
\end{equation}
with $\omega_k=\sqrt{k^2+m_\pi^2}$. 
For the total number of pions of all species ($\pi^+,\pi^0,\pi^-$) we
have: 
\begin{equation}
\frac{N_{tot}(t)}{V}=\sum^3_{i}\int\frac{d^3k}{(2\pi)^3}\,n^{}_{ki}(t).
\label{nol}
\end{equation}
where $V$ is the volume of our system, that is the volume of the 
fireball in a heavy-ion collision experiment.

\section{Initial Conditions}

Our choice of the vacuum at $v$ as our reference state has 
the advantage that
the particle interpretation of the field $\vp(\vx,t)$ is close to the
experimentally observable quantities. It requires, however, some care with
respect to our choice of initial conditions for the evolution described by
equations (\ref{eqs}), (\ref{eqp}). 
We assume that the fireball created by the collision
is initially in local thermodynamic equilibrium, 
or it has been separated in Disoriented Chiral Condensates
(DCCs), each one in its own local thermodynamic equilibrium. 
If the second case is realised,
our treatment applies to the interior of one of these DCCs.
The initial expectation value $\s_1$ that we consider for
the $\sigma$-field is small, but non-zero, because
of the explicit chiral symmetry breaking.
For our calculation we use $\sigma_1=0.1 v$ at $T=140$ MeV.

For the pions we expect initially 
a thermalised gas that follows a Bose-Einstein
distribution with
\begin{equation}
n^{\rm{eq}}_{ki}=\frac{1}{e^{\frac{\omega_k}{T}}-1}.
\label{nkieq}
\end{equation}
We assume the dispersion relation around the vacuum at $\s=v$: 
$\omega^2_k=k^2+m_\pi^2$, even though the pion mass depends on 
the temperature. The justification for this approximation is
provided by the explicit study in \cite{BJW} of the effective pion mass during 
the process of chiral restoration. There, it is shown that $m_\pi$ stays 
approximately constant from $T=0$ up to $T\sim 100$ MeV.

The mode functions $f_{ki}(t)$, for a configuration corresponding to
a non-interacting pion gas in thermal equilibrium, can be taken
\begin{equation}
f^{\rm{eq}}_{ki}=\sqrt{n^{\rm{eq}}_{ki}+1/2}\,\,\,\frac{e^{-i\omega_kt}}
{\sqrt{\omega_k}},
\label{fkeq} 
\end{equation}
in agreement with (\ref{nki}).
In the following we restrict our analysis to large occupation numbers, and 
neglect the factor $1/2$ related to
the zero point energy. 

In our simplified scenario we assume an instantaneous 
expansion of the fireball by a
volume factor $\Lambda$ (a quench). 
This means that the number densities of the pion gas must be reduced by 
the same factor. 
In addition, in order to be consistent with the conservation of energy, 
the initial value of the
$\s$-field has to change according to the relation
\begin{equation}
V(\s_\Lambda)=\frac{V(\s_1)}{\Lambda},
\label{vsl}
\end{equation}
where $\s_1$ is the value before the quench, and 
$\s_\Lambda$ the one after.
We point out that this assumption is rather crude as it neglects possible
fluctuations of $\s$. However, it satisfies the minimal requirement of 
energy conservation.

The above discussion implies that 
the physically motivated initial conditions for the evolution 
of the fields are 
\begin{equation}
f_{ki}(0)=\sqrt{\frac{n^{\rm{eq}}_{ki}}{\Lambda}} \, 
\frac{1}{\sqrt{\omega_k}}\  ,
\ \ \
\dot{f}_{ki}(0)=\sqrt{\frac{n^{\rm{eq}}_{ki}}{\Lambda}} \, 
\left(-i\sqrt{\omega_k}\right)
\label{fki0}
\end{equation}
and
\begin{equation}
\s(0)=\s_\Lambda \ , \ \ \
\dot{\s}(0)=0.
\label{s0}
\end{equation}
These initial conditions differ from the ones assumed 
for particle production through inflaton decay in cosmology
\cite{LIC}, and in some works on QCD \cite{K,DS}, 
as in those cases the initial particle number is taken to be zero.

\section{Non-Equilibrium Evolution} 

Equations (\ref{eqs}) and (\ref{eqp}), determining the evolution of the $\s$
and the pion field, constitute a non-linear integro-differential system 
that is not solvable analytically. The main difficulty is related to 
the presence of the  term
$\langle\vp^2(t)\rangle$ that mixes all
the pion mode functions.
If we ignore this term, it is possible to decouple the evolution of $\s$
from that of the pion field. We introduce the variable 
 $\n(t)={\s(t)}/{v}$, 
and absorb
$\la^2v^2$ in a new rescaled time variable 
$\tau=\la vt$.
We also introduce the small parameter 
$\e={m_\pi^2}/({2\la^2v^2})\ll 1$. In this way we obtain
\begin{equation}
\n''(\tau)+\n^3(\tau)+(2\e-1)\n(\tau)-2\e=0,
\label{eqn0}
\end{equation}
where primes denote differentiation with respect to $\tau$.

In zeroth order in $\e$, i.e $\e=0$, this equation can be solved
in terms of elliptic functions, with the result
\begin{equation}
\n(\tau)=\frac{\n(0)}{{\rm dn}\left( \tau\sqrt{1-\frac{n^2(0)}{2}},q\right) },
\label{ntau}
\end{equation}
where
\begin{equation}
q=\sqrt{\frac{1-\n^2(0)}{1-\frac{\n^2(0)}{2}}}.
\label{modulus}
\end{equation}
Here dn stands for the known Jacobi elliptic function and $q$ is its modulus.
(We do not use the standard symbol $k$ in order to avoid confusion with 
momentum.)
We point out that the 
solution (\ref{ntau}) holds only for the range $0\leq\n(0)\leq1$, 
relevant to our case.
For $\n(0)>1$
the solution is expressed in terms of different elliptic functions.

Inserting the solution (\ref{ntau}) into the equation 
for the pion mode functions, we obtain
\begin{equation}f''_{ki}(\tau)+\left[\frac{k^2}{\la^2v^2}
+\frac{\nos}{{\rm dn}^2\left( \tau\sqrt{1-\frac{\nos}{2}},q \right)}
+2\e-1\right] f_{ki}(\tau)=0.
\label{eqf0}
\end{equation}
This is the Lam\'e equation, for particular values of the coefficients 
which make it solvable in terms of Jacobi functions \cite{INCE,RZ}.

We first investigate the zeroth order solution ($\e=0$). 
As we show in the Appendix A, the solution $f_{ki}(\tau)$ of 
(\ref{eqf0}) is a quasi-periodic function that can grow 
with time. Exponential amplification is obtained if the pion momentum
lies in one of the 
zones: a) $-\infty\leq{k^2}/({\la^2v^2})\leq0$, 
~b) ${\nos}/{2}\leq{k^2}/({\la^2v^2})\leq1-{\nos}/{2}$.
In our case the first zone is unphysical.
Substituting $\s(t)=v\n(t)$, we obtain for the remaining amplification zone
\begin{equation}
\frac{\la}{\sqrt{2}}\s(0)\leq k\leq\frac{\la}{\sqrt{2}}\sqrt{2v^2-\s^2(0)},
\label{zone0}
\end{equation}
a result valid in zeroth order in $\e={m_\pi^2}/({2\la^2v^2})$.

Obtaining a full analytical solution valid 
in first order in $\e$ is very complicated.
However, it is trivial to take into account the term $2 \e$
in equation (\ref{eqf0}). Its consideration simply leads to the replacement of
${k^2}/({\la^2v^2})$ by 
$2\e+{k^2}/({\la^2v^2})$.
The amplification zone becomes
  \begin{equation}
\sqrt{\frac{\la^2}{2}\s^2(0)-m_\pi^2}\leq k\leq
\sqrt{\frac{\la^2}{2}(2v^2-\s^2(0))-m_\pi^2}.
\label{zonee}
\end{equation}
We emphasize that this result is not the complete answer in first order
in $\e$, as equation (\ref{ntau}) is not an exact solution of (\ref{eqn0}) for
$\e\not=0$.
However, the amplification zone given by equation (\ref{zonee})
agrees very well (within 4\%)
with the amplification observed through the numerical solution of
the equations.

If $\s(0)$ is sufficiently small 
($\s(0)<\sqrt{\frac{2}{\la^2}}m_\pi\simeq0.5\,v$) 
the lower end of the amplification zone (\ref{zonee}) is at $k=0$. 
For larger $\s(0)$ the zone becomes narrower. 
In the extreme case $\s(0) \simeq v$
(when the $\s$-field starts its evolution very close to its minimum),
the zone shrinks to a 
point
\begin{displaymath}
k^2\simeq\frac{m_\s^2}{4}-m_\pi^2.
\end{displaymath}
This is the pion momentum in the decay $\s\rightarrow2\pi$.

The form of 
equation (\ref{eqp}) for the pionic modes
indicates the presence of two regimes in their evolution:\\
a) Shortly after the quench the various modes have an effective
mass term
\begin{equation}
m^2_{eff}=
k^2-\la^2v^2+\la^2\s^2(0)
+\frac{5}{3}\la^2\langle\vp^2(0)\rangle+m_\pi^2.
\label{effmass}
\end{equation}
One expects the exponential growth of the low-momentum
modes for which the mass term is negative.
This phenomenon is characterized as spinodal decomposition \cite{Br,BLS}.\\
b) At later stages 
the mass term becomes
\begin{equation}
m^2_{eff}(t)=
k^2-\la^2v^2+\la^2\s^2(t)
+\frac{5}{3}\la^2\langle\vp^2(t)\rangle+m_\pi^2,
\label{effmasst}
\end{equation}
with a $\s(t)$ a quasi-periodic function.
If the term $\sim \langle\vp^2(t)\rangle$ is negligible the evolution equation becomes a Lam\'e or Mathieu equation
(see Appendix B for a detailed discussion of this point and
a related controversy). One expects a resonance band with 
exponential amplification of $f_k$. This phenomenon is characterized
as parametric resonance. It has been studied in detail in relation to
the reheating of the universe through inflaton decay \cite{KLS,B,LIC,STB}.

The analytical solution that we presented above, as well as the
numerical analysis in the next section, demonstrate that there is
no well-defined boundary separating periods in the time evolution during
which one of the two mechanisms dominates. In general, there is a fast
amplitude growth for the modes in a certain momentum zone (such as in
equation (\ref{zonee})). 
However, the phenomenon can be a complicated convolution
of both mechanisms we discussed above.

\section{Numerical Results and Discussion}

We solve equations (\ref{eqs}) and (\ref{eqp}) numerically using a 
fourth-order Runge-Kutta algorithm for the differential 
equations and an 11-point Newton-Gotes integrator to compute the
momentum 
integral:
\begin{equation}
\langle\vp^2(t)\rangle=\int\frac{d^3k}{(2\pi)^3}\vf^\star(t)\vf(t)=
\int\frac{dk}{2\pi^2}\,k^2\vf^\star(t)\vf(t).
\label{pi2tnum}
\end{equation}
We calculate the pion density in $3D$ momentum space using (\ref{nki})
and the total number of produced pions $N_{tot}$ using (\ref{nol}).
Furthermore, it is convenient to define the projected $1D$ density 
$\rho(k)$ through:  
\begin{equation}
N_{tot}=\int_0^\infty dk \rho(k).
\label{rho}
\end{equation}
The pion gas is initially a fireball at thermal equilibrium with temperature 
$T \simeq 140$ MeV and radius $r_0 \simeq 10$ fm. The quench is described
by an instantaneous expansion of this fireball by a factor $\Lambda$.
We consider here three different values of the final radius corresponding to small ($r_f=11$fm ($\Lambda=1.3$)), intermediate ($r_f=12.6$ fm ($\Lambda=2$)) and large ($r_f=15$fm ($\Lambda=3.375$)) expansion of the initial fireball. 
The effect of the quench is incorporated in the normalization  
of the pion mode functions $\vf$ at $t=0$. Performing the numerical integration
we obtain the results presented in Figs.~1--8, which are close to our analytical
calculations described in section IV. 

In Fig.~1 we depict the $\s$ evolution and the total pion number for expansion factor 
$\Lambda=3.375$. The radius of the expanded fireball in this case is $15$ fm. During the first 
two oscillations of $\s(t)$ the number of produced pions $N_{tot}$ increases fast, because of
parametric resonance. This lasts about $6$ fm and subsequently $N_{tot}$ just fluctuates
around a mean value $\simeq 830$. As our approach is semiclassical
this solution describes the evolution of the system consistently only 
within the first stage
($t \lta 6$ fm). During this time the pion production can be
described through the energy transfer from the oscillating $\sigma$-field 
to the pionic field. In the following stages, our solution displays a periodic
energy exchange between the two fields. 
A proper quantum treatment is necessary in order to describe the complete
decay of the $\sigma$-field into free pions. However, the bulk of pions is
produced during the first stage. 
For this reason we estimate the various observables (such as pion number and
distribution)
at a time $t \simeq 6$ fm.    

\begin{figure}
\includegraphics{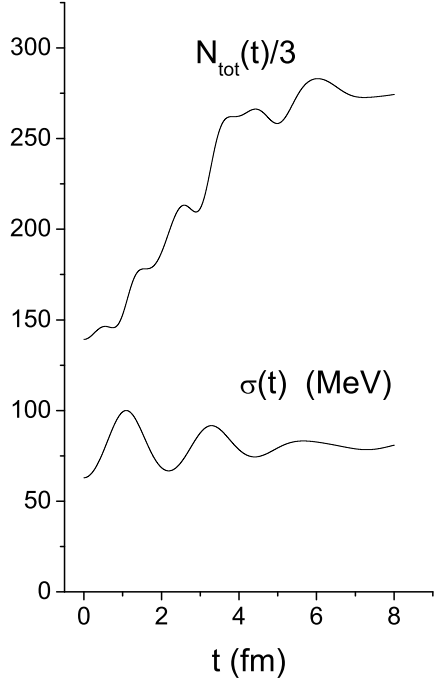}
\caption{\label{fig:fig1} The evolution of the $\s$-field as well as the total pion number 
$N_{tot}(t)$ (scaled by a factor of 3) for initial 
radius $10$ fm and final radius $15$ fm ($\Lambda=3.375$). }
\end{figure}

In Fig.~2 the evolution of the distribution of pions $\rho(k)$ for
various times is shown. We observe a large enhancement 
of the spectrum at low momenta, and the formation of a zone with a
peak at a specific $k$-value ($k_{max}$) and a width characteristic
of the non-equilibrium amplification. Comparing with Fig.~1 we observe that
the energy transfer to the pions is translated to a decrease of the 
$\sigma$-oscillation amplitude. 
In Fig.~2 we also observe a 
slight  sinking of the spectrum at large momenta,
which implies 
an additional energy transfer from hard to soft pion modes through
the mode-mode coupling (the $\langle \vec{\pi}^2(t)\rangle$ term) in (6), effect that cannot be
described within our approximations.

\begin{figure}
\includegraphics{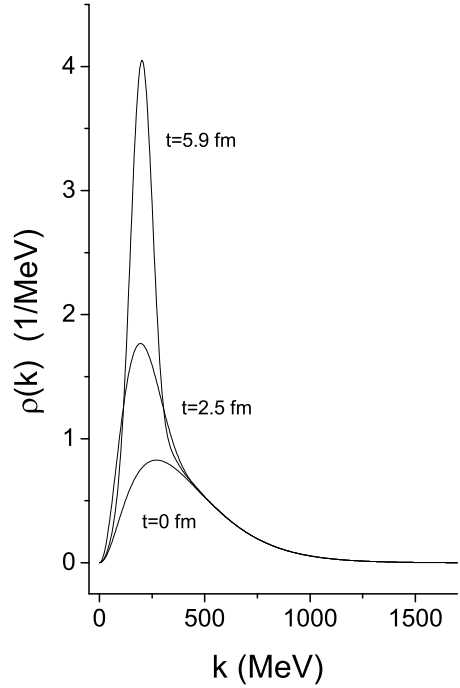}
\caption{\label{fig:fig2} The density $\rho(k)$ defined in equation
(\ref{rho}) using 
$\Lambda=3.375$ at three successive times.}
\end{figure}

This behaviour can be described very well using the analytical results 
of section IV.
For an expansion factor $\Lambda=3.375$ (and more generally for $\Lambda>3$)
the initial values $f_{ki}(0)$ (according to (\ref{fki0})) are very
small so that $\langle\vp^2(0)\rangle\ll v^2$, justifying the conditions for
parametric resonance: while $\s$ oscillates around its new minimum after the
quench, it leads 
the pion mode functions in the momentum zone
(\ref{zonee}) to exponential increase. The numerical values for the 
limits of the momentum zone, whithin which the exponential amplification occurs,
agree  with a $4\%$ accuracy with the analytical estimate of equation
(\ref{zonee}). This is explained by the fact that $\langle\vp^2(t)\rangle$ is negligible initially in this case, 
so that the analytical solution gives a good approximation. As time passes, 
$\langle\vp^2(t)\rangle$ increases, so that eventually 
the parametric resonance ends. 
Even though this effect cannot be 
described within the analytical approach, its influence on the location of the amplification zone is very small. 

We also point out that the initial conditions for $\s(t)$ are determined
by equation (\ref{vsl}), 
so that a large $\Lambda$ corresponds to $\s(0)$ closer to
$v$. This large initial value of 
$\sigma$ shifts the pion effective mass at
$t=0$ to larger values, leading to a strong suppression of spinodal
decomposition in favour of parametric resonance. 

In Fig.~3 we plot $\s(t)$ and $N_{tot}(t)$ for $\Lambda=1.3$. We observe 
that the pion production completes during the rolling down of the $\sigma$-field 
towards the new minimum. This process lasts for $\simeq 2$ fm, and
subsequently $N_{tot}$ oscillates 
in a manner similar to that in the case $\Lambda=3.375$. 
Observables are now considered for times
$t\simeq 2$ fm. 
For small expansion factors ($\Lambda<1.5$) the
dominant amplification mechanism is the spinodal decomposition: during the 
first
rolling down of the $\sigma$-field towards its minimum, the effective squared pion mass (\ref{effmass}) is
negative and the pion modes with small momenta are amplified. 
Furthermore, because of the large  value of $\langle\vp^2(0)\rangle$,
the requirements of parametric resonance are not fulfilled in the second stage
when $\s(t)$
oscillates around the new minimum.
As a result the amplification through parametric resonance is suppressed.
We point out that the analytical solution of the previous section does not give a good quantitative approximation in this case, as
the effect of the term $\sim\langle\vp^2(t)\rangle$ cannot be neglected.

\begin{figure}
\includegraphics{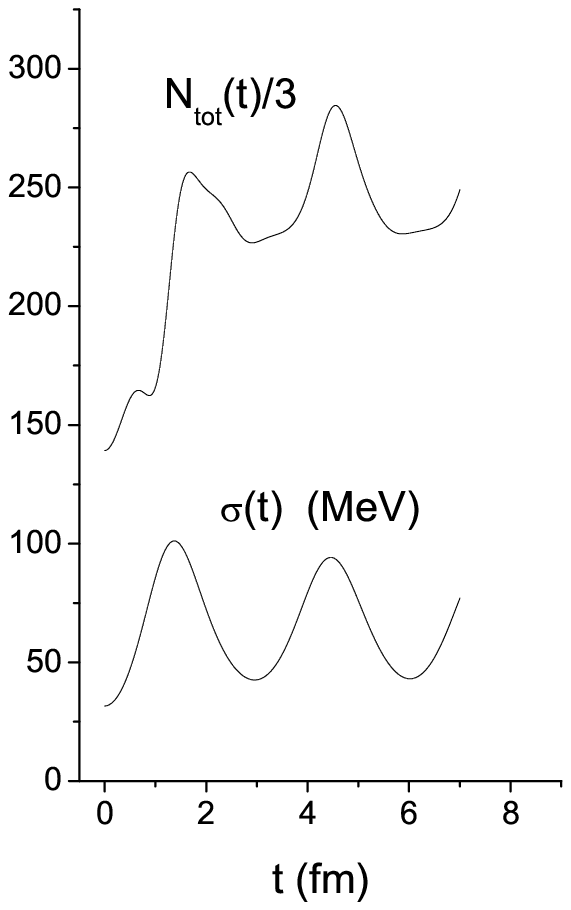}
\caption{\label{fig:fig3} The evolution of the $\s$-field as well as the total pion number 
$N_{tot}(t)$ (scaled by a factor of 3) for expansion factor $\Lambda=1.3$. }
\end{figure}

For completeness, we show in Fig.~4 the evolution of the pion
distribution for $\Lambda=1.3$. Similarly to the previous case (Fig.~2) we observe an
enhancement of the pion spectrum at low momenta.
The lower limit of the ampilified zone is at $k=0$, while the upper limit
is time dependent. The peak of the particle density per momentum mode
is at $k=0$. However this feature is 
modified at the level of one-particle momentum density (\ref{rho}), 
because of the phase space 
factor 
$4\pi k^2$ that shifts the maximum to a non-zero momentum value. 

\begin{figure}
\includegraphics{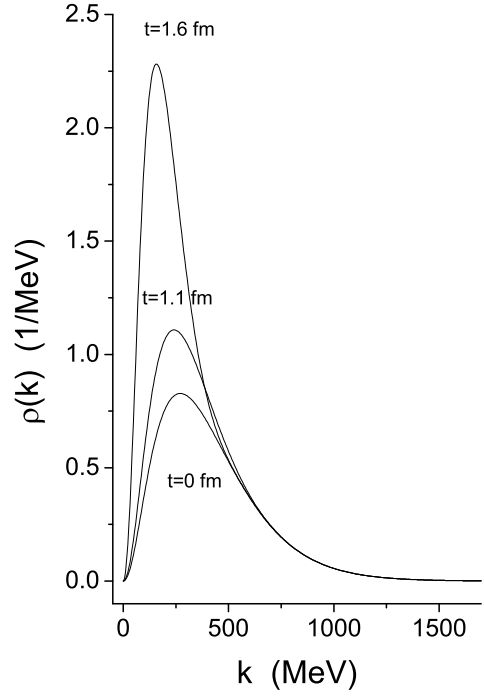}
\caption{\label{fig:fig4} The density $\rho(k)$ using 
$\Lambda=1.3$ at three successive times.}
\end{figure}

Lastly, for intermediate expansion factors ($1.5<\Lambda<3$) both amplification
mechanisms are operative. 
During the first $\s$-rolling, 
$m_{eff}(t)$ is negative for low momenta and spinodal 
decomposition takes place. Subsequently, while $\s$ oscillates around its new minimum, 
$\langle\vp^2(t)\rangle$ is not very large, allowing for parametric resonance to occur as
well. This behaviour can be clearly seen in Fig.~5: the pion production happens during both 
phases. At late times $\langle\vp^2(t)\rangle$ grows significantly, parametric resonance 
ends, and pion production completes. The time when this happens is $\simeq~ 4$
fm.

\begin{figure}
\includegraphics{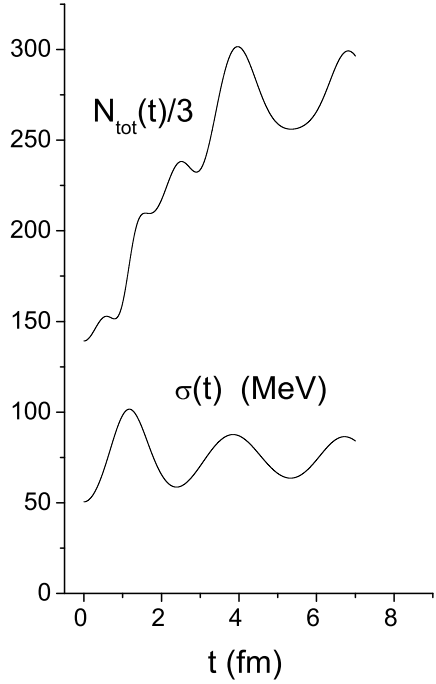}
\caption{\label{fig:fig5} The function $\s(t)$ as well as the total pion number 
$N_{tot}(t)$ (scaled by a factor of 3) for an expansion factor $\Lambda=2$. }
\end{figure}

In this intermediate case both limits of the amplification zone are
time dependent. As a result, the pion momentum spectrum changes with time. 
In fact, we get an
oscillating zone in momentum space within which the amplification takes place. 
A simple
explanation of this behaviour can be obtained 
if we observe that the addition of 
a constant 
term within the brackets in equation (\ref{eqf0}) 
shifts the amplification zone given by equation (\ref{zonee}), to the left if the
added term is positive, or to the right if it is negative. 
The addition of a 
fluctuating term ($\langle\vp^2(t)\rangle$) is expected to force the edges of the amplification zone to
oscillate. 
It must be pointed out that, after the 
inclusion of the back-reaction term $\langle\vp^2(t)\rangle$ in the 
equation for $\sigma$, the solution 
(\ref{ntau}) is no longer valid.
However, in this intermediate case 
the modification of the solution is not expected to be 
dramatic. 
The expected behaviour is observed in Fig.~6.
The pion spectrum is enhanced and 
its maximum is shifted towards lower momenta as in the previous two cases.
However, the location of the maximum of the spectrum 
and the limits of the amplification zone oscillate.

\begin{figure}
\includegraphics{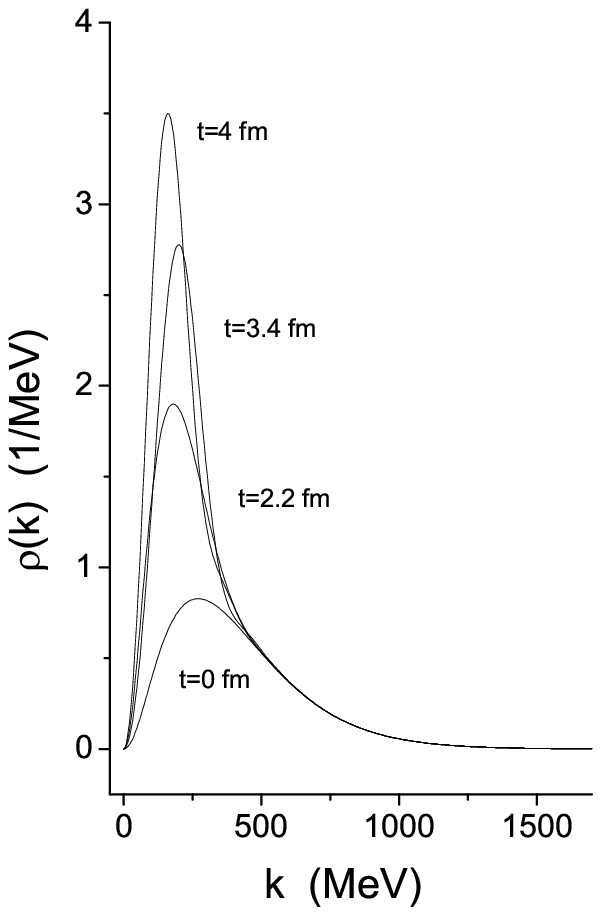}
\caption{\label{fig:fig6} The density $\rho(k)$ for the case of intermediate expansion 
($\Lambda=2$) at four successive times.}
\end{figure}

A common property of the momentum spectrum of the produced pions, 
for all the expansion factors we considered, 
is the appearance of an amplification zone in the region of low momenta. 
This results in the production of a relatively large number of soft pions 
during the evolution 
of the $\sigma$-$\pi$ system. 
As we have already mentioned, our semiclassical approach is not reliable 
for long times.
We estimate the number of produced pions through the pion distribution
at the beginning of the stage in which our solution 
predicts a periodic interchange of 
energy between
the $\sigma$-field and pion sectors: $\simeq 6$ fm for 
$\Lambda=3.375$, $\simeq 4$ fm for $\Lambda=2$ 
and $\simeq 2$ fm for $\Lambda=1.3$. 
The effect of the amplification is quite strong, 
and persists even if we perform an averaging of the solution over times 
larger than the ones listed above.
The energy density 
$\rho(k)$ is clearly distinguishable from that of a 
thermal system because of the presence of a narrow peak at low momenta.

The formation of the momentum amplification zone reflects a clustering of 
the pion momenta 
in the low $k$-region. The dynamical character of this cluster, 
which leads to the presence
of a peak at a definite momentum value, accompanied by a width of 
typical size, in the 
projected one-particle density, is crucial. This characteristic form can be
directly observed in an inclusive analysis of the pion momenta
produced for example in $A+A$ ultrarelativistic collisions. 
The geometry of 
this cluster is expected to be rather simple: no substructures at 
smaller scales 
are present. It is worth comparing the situation here with the corresponding 
formation of 
particle clusters in a critical system. In the latter case the clusters have 
statistical
character. Therefore, at the level of one-particle inclusive density
in momentum space no peaks (beyond the kinematic ones) appear. The geometry 
of these critical 
clusters is that of a random fractal: 
self-similar structures at different scales 
occur, leading to a characteristic power-law dependence of the 
factorial moments on the
resolution scale $M$ \cite{ACDPRE}. 

In order to explore the resonant pion production at a phenomenological level, 
we have generated,
through a Monte-Carlo simulation of 
equations (\ref{nol}),(\ref{rho}),
data sets consisting of a large number of events. 
For each value of the expansion factor $\Lambda$ that we considered, 
we produce 1000 events. We then calculated the corresponding 
second factorial moment in 
transverse momentum space as a function of the resolution scale 
$M$ \cite{BialPes}. Our results
are given in Fig.~7. We observe a conventional behaviour of the 
moments (saturation at high 
resolution scales) reflecting the absence of self-similar structure at 
different scales in the 
formed pionic clusters, for all expansion factors.

\begin{figure}
\includegraphics{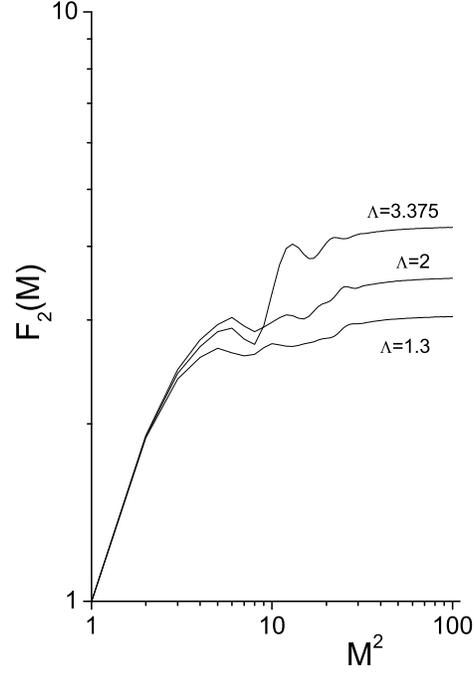}
\caption{\label{fig:fig7} The second factorial moment for the three different cases of 
large ($\Lambda=3.375$), intermediate ($\Lambda=2$) and small ($\Lambda=1.3$) expansion.
For each case we have analysed 1000 Monte-Carlo events generated according to 
equations (\ref{nol}), (\ref{rho}).}
\end{figure}

One could also consider 
the inclusive density of the transverse momentum for each set of 1000 events. 
We have calculated
this distribution for the case $\Lambda=3.375$. The result is displayed 
in Fig.~8, where we
clearly observe the effect of the amplification zone. In the same plot
we show the experimentally observed transverse momentum distribution 
of the charged pions
produced in $Pb+Pb$ collisions at $158$ GeV/n (NA49-SPS) 
\cite{Alber:1995ws}.
The experimental results are very close to a thermal system in transverse 
momentum space. The large deviation of the transverse 
momentum distribution of the pions produced through 
the amplification mechanisms described in 
the present work, from the corresponding distribution of a conventional 
thermal system, is an
experimentally accessible signal for the appearance of 
non-equlibrium phenomena, associated with chiral phase transition, in the pion
production during a heavy-ion collision. 

\begin{figure}
\includegraphics{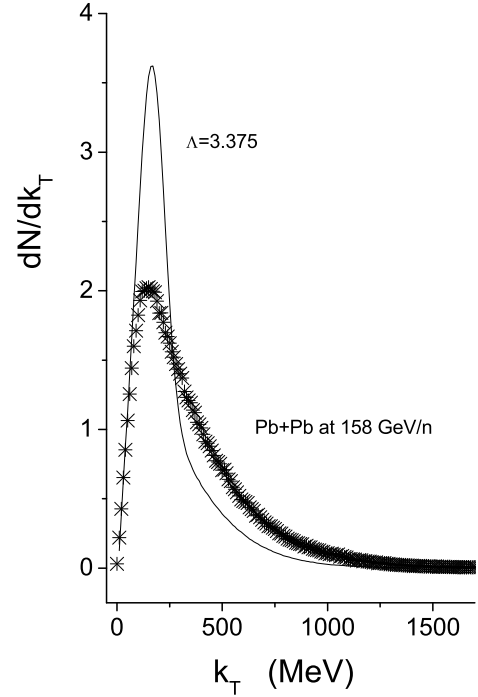}
\caption{\label{fig:fig8} The transverse momentum distribution of the finally produced pions
in the case of large expansion ($\Lambda=3.375$). For comparison the $k_T$-distribution for
a system at thermal equilibrium ($Pb+Pb$ at $158$ GeV/n) measured at the NA49 experiment 
(CERN-SPS) is also displayed. Both distributions are normalized to the same total pion
multiplicity.}
\end{figure}

\section{Conclusions}

In this work we studied non-equlibrium phenomena related
to the chiral QCD phase transition. We found that, within the scenario
of quench, the dynamics of the system may amplify pion modes in a 
certain momentum zone while
the $\sigma$-field moves towards and around the (chirally broken) 
vacuum.
These phenomena produce a significant enhancement of the pion spectrum at 
low momenta. In our 
approach we took account of two main features of the physical system:
the explicit symmetry breaking 
through the non-zero pion mass, and 
the local thermal equilibrium for the initial pion gas. 
In this sense our study is close to the real world. 
As a consequence, the pion production in our model 
is not as pronounced as in other models \cite{K,DS,B}. 
However, the momentum distribution of the
produced pions possesses fingerprints of the zone structure 
of the amplification regions in 
momentum space: a cluster of pions with low momenta is formed. 
Although visible at the
level of single particle density, this cluster does not have substructures 
at different scales. It
is, therefore, distinguishable from particle clusters created by different 
mechanisms 
(equilibrium phase transition). Furthermore, the transverse momentum spectrum
acquires a specific non-thermal peak that could be a clear signature of 
these non-equilibrium 
phenomena in a collision experiment with relativistic heavy ions.

It would be interesting to examine the modifications in the scenario we studied,
if the finite time 
neccessary for the quench is taken into account. 
Additionally, one should also consider the 
consequences of the cylindrical form of the expanding fireball 
for the observed pion 
spectrum. 
Such a geometry is closer to the conditions occuring in $A+A$ 
experiments at very high energies. Another important aspect is the case 
when the initial state is
critical, i.e. large self-similar fluctuations of the $\sigma$-field 
are present. In this case, the interesting  
problem is to explore the evolution of the critical state 
towards the freeze-out phase. 
The study of these additional issues is left for future investigations. 

\section*{Appendix A}

We present here the derivation of some of the results used in the
main text.
Equation (\ref{eqf0}) with $\e=0$ has the form
\begin{equation}
f''_{ki}(\tau)+\left[\frac{k^2}{\la^2v^2}
+\n^2(\tau)-1\right]
f_{ki}(\tau)=0
\label{eqf0app}
\end{equation}
with
\begin{equation}
\n(\tau)=\frac{\n(0)}{{\rm dn}\left(
\tau\sqrt{1-\frac{n^2(0)}{2}},q\right) },
\label{ntau1}
\end{equation}
\begin{equation}
q=\sqrt{\frac{1-\n^2(0)}{1-\frac{\n^2(0)}{2}}}.
\label{modulusa}
\end{equation}
Here $\n(\tau)$ is a doubly periodic function  with periods
 $2\om={2K(q)}/{\sqrt{1-\frac{\nos}{2}}}$
(real) and  $4\op={4iK'(q)}/{\sqrt{1-\frac{\nos}{2}}}$
(complex), where $K(q)$ is the complete elliptic integral of first
kind and 
$K'(q)$ is its complementary.

Equation (\ref{eqf0app}) is the Lam\'e equation, 
for a particular value of the coefficients 
that make it solvable in terms of Jacobi functions.
It is convenient
to express the Jacobi function dn in terms of the Weierstrass 
$\mathcal{P}$ function through \cite{RZ}
\begin{equation}
\frac{1}{{\rm dn}^2(u,q)}=\frac{1-\frac{1}{{\rm
sn}^2(u+K(q)+iK'(q))}}{1-q^2}=
\frac{1}{e_1-e_2}\left[e_1-\mathcal{P}\left(\frac{u+K(q)+iK'(q)}
{\sqrt{e_1-e_3}}\right)\right],
\label{dnweis}
\end{equation}
where $q=\sqrt{\frac{e_2-e_3}{e_1-e_3}}$. 
The parameters $e_1,e_2,e_3$ are the roots of a
cubic equation whose coefficients depend on the two periods 
of the Weierstrass 
$\mathcal{P}$ function \cite{RZ}.
They obey the relation $e_1+e_2+e_3=0$. Without loss of
generality we choose
$e_1-e_3=1-\frac{\nos}{2}$. Thus we obtain
\begin{eqnarray}
 &3e_1=1\nonumber\\ 
&3e_2=1-\frac{3}{2}\nos\nonumber\\ 
&3e_3=\frac{3}{2}\nos-2.
\label{roots}
\end{eqnarray}

Inserting (\ref{dnweis}) into (\ref{eqf0app}) and using the
definitions 
of the periods $\om,\op$, we find
\begin{equation}
f''_{ki}(\tau)+\left[\frac{k^2}{\la^2v^2}-\frac{1}{3}
-2\mathcal{P}(\tau+\om+\op)\right]\, f_{ki}(\tau)=0.
\label{lameweis}
\end{equation}
 The solutions of (\ref{lameweis}) are of the form \cite{INCE}
\begin{equation}
f_{ki}(\tau)=\frac{\s(\tau+\om+\op+\w(k))}{\s(\tau+\om+\op)}\, 
e^{-\tau\zeta(\rm{w}(k))},
\label{fkitau}
\end{equation}
where $\s(x)$ and $\zeta(x)$ are the known Weierstrass functions
defined as
\begin{equation}
\zeta(u)=\frac{1}{u}-\int_0^u\left(\mathcal{P}(z)-\frac{1}{z^2}\right)dz
\end{equation}
and 
\begin{equation}
\s(u)=u\exp\left\{\int_0^u\left(\zeta(z)-\frac{1}{z}\right)dz\right\},
\end{equation}
 and the complex number $\w(k)$
 is defined through the transcendental equation 
\begin{equation}
\mathcal{P}(\w)=-\frac{k^2}{\la^2v^2}+\frac{1}{3}.
\label{wdef}
\end{equation}

Because of the quasi-periodicity property of the $\s$-function
\cite{RZ}
\begin{displaymath}
\s(x+2\om)=-\s(x)\,e^{2(x+\om)\zeta(\om)},
\end{displaymath}
we obtain
from (\ref{fkitau}) 
\begin{equation}
f_{ki}(\tau+2\om)=f_{ki}(\tau)\,e^{2\left[\w(k)\,\zeta(\om)-\om\zeta(\w(k))
\right]}.
\label{fkitauom}
\end{equation}
As a result, for $\w(k)$ for which the 
exponent
\begin{equation}
\mu(\w(k))=2\left[\w(k)\zeta(\om)-\om\zeta(\w(k))\right]
\label{expo}
\end{equation}
becomes
real, we have an exponential amplification of $f_{ki}(\tau)$. For
values of
$\w(k)$ 
which lead to imaginary $\mu(\w(k))$ we obtain oscillatory 
behaviour.

To determine the corresponding $\mu(\w(k))$ intervals one has to
consider
the equation (\ref{wdef}) that maps the real ${k^2}/({\la^2v^2})$ 
axis onto the sides of the 
fundamental square in the w plane \cite{B,RZ}.
\begin{itemize}
\item{If $-\infty\leq\frac{k^2}{\la^2v^2}\leq0$, then $\w(k)=\beta\in
R$ with 
$0\leq\beta\leq\om$.\ \ \ \ \ \ \ \ \ \ \ \ \ \ \ \ \ \ \ \ \ \ \ \ \ \ \
\ \ \ \ (A)}
\item{If $0\leq\frac{k^2}{\la^2v^2}\leq\frac{\nos}{2}$, then 
$\w(k)=\om+i\alpha$ 
 $(\alpha\in R)$ with $0\leq\alpha\leq-i\op$.\ \ \ \ \ \ \ \ \ \ \ \ \ \
(B)}
\item{If
$\frac{\nos}{2}\leq\frac{k^2}{\la^2v^2}\leq1-\frac{\nos}{2}$, then 
$\w(k)=\op+i\beta$ with $\om\geq\beta\geq0$.\ \ \ \ \ \ \ \ \ \ \ \ \ \ \ 
\ \ \ \ (C)}
\item{If $1-\frac{\nos}{2}\leq\frac{k^2}{\la^2v^2}\leq\infty$, then 
$\w(k)=i\alpha$ with $-i\op\geq\alpha\geq0$.\ \ \ \ \ \ \  \  \ \ \ \ 
\ \ \ \ \ \ \ \ \ \ \ \ (D)}
\end{itemize}
The appearance of $\n(0)$ in these four intervals 
results from the dependence of the roots $e_1,e_2,e_3$ in (\ref{roots})
on $\n(0)$.
Using the properties of the 
$\zeta$-function \cite{RZ} we can easily derive that 
$\mu(\w(k))$ has a non-zero real positive 
part for the cases (A) and (C), while it is purely
imaginary for (B) and (D).

\section*{Appendix B}

We derive the Mathieu equation starting from the 
Lam\'e equation.
We use the expansion \cite{RZ} 
\begin{equation}
\frac{1}{{\rm
dn}(u,q)}=\frac{\pi}{2q'K(q)}\left[1+4\sum^\infty_{n=1}(-1)^n
\frac{Q^n}{1+Q^{2n}}\,\cos\frac{n\pi u}{K(q)}\right],
\label{dnexpa}
\end{equation}
 where
\begin{displaymath}
Q\equiv e^{-\frac{\pi K'(q)}{K(q)}}
\end{displaymath}
is the elliptic nome. This expansion holds for 
${\rm Im}\left(\frac{\pi u}{2K(q)}\right)<\frac{1}{2}\pi\,
{\rm Im}\left(\frac{i K'(q)}{K(q)}\right)$ \footnote{This is always
satisfied in our model, 
since $q=\sqrt{\frac{1-\nos}{1-\frac{\nos}{2}}}$ with
 $0<\n(0)<1$, so that $q$, $K(q)$, $K'(q)$ are always real.}.
$Q$ can be expanded as \cite{RZ} 
\begin{equation}
Q=e^{-\frac{\pi
K'(q)}{K(q)}}=\lambda+2\lambda^5+15\lambda^9+150\lambda^{13}+\cdots
\label{qexpa}
\end{equation}
where
\begin{displaymath}
\lambda=\frac{1}{2}\,\frac{1-\sqrt{q'}}{1+\sqrt{q'}}=
\frac{1}{2}\,\frac{1-(1-q^2)^{1/4}}{1+(1-q^2)^{1/4}}.
\end{displaymath}
$K(q)$ can be also expanded as
\begin{equation}
K(q)=\frac{\pi}{2}\,
\sum_{n=1}^\infty\left(\frac{(2n-1)!!}{2^nn!}\right)^2q^{2n}.
\label{kexpa}
\end{equation}
Keeping only the first terms in the expansions 
we obtain
\begin{equation}
\n^2(\tau)= 
\frac{\n^2(0)}{\rm{dn}^2\left(\tau\sqrt{1-\frac{n^2(0)}{2}},q\right)}=
\frac{\nos\pi}{2(1-\frac{1}{3}q^2-\frac{7}{64}q^4)}
\left[1-8(\lambda+2\lambda^5)\,\cos\left(\frac{2\tau\sqrt{1-\frac{n^2(0)}{2}}}
{1+\frac{1}{4}q^2+\frac{9}{64}q^4}\right)\right]
\label{nexpa}
\end{equation}
with $q=\sqrt{\frac{1-\nos}{1-\frac{\nos}{2}}}$.
Substituting this in the Lam\'e equation (\ref{eqf0app})
we find
\begin{equation}
\left[\frac{d^2}{dz^2}+A-2B\cos2z\right]f_{ki}(z)=0
\label{mathieu}
\end{equation}
with 
\begin{eqnarray}
A&=&\frac{k^2}{\la^2v^2}+2\e-1+
\frac{\nos\pi}{2(1-\frac{1}{3}q^2-\frac{7}{64}q^4)}\nonumber\\
B&=&2\,\frac{\nos\pi}{(1-\frac{1}{3}q^2-\frac{7}{64}q^4)}\,
(\lambda+2\lambda^5)\nonumber \\
z&=&\frac{\tau\sqrt{1-\frac{n^2(0)}{2}}}
{1+\frac{1}{4}q^2+\frac{9}{64}q^4}.
\label{matpara}
\end{eqnarray}
Equation (\ref{mathieu}) is the Mathieu equation. It is a good 
approximation of the Lam\'e equation for $Q\ll 1$, i.e. for $q\ll 1$, 
since the elliptic functions approach the usual trigonometric ones
for $q\rightarrow 0$. However, 
in the case $Q~\sim~1$, i.e for $q~\sim~1$, 
all higher-order terms in (\ref{dnexpa}) must be taken into account.
The simple approximation that leads to the
Mathieu equation is not valid. There is some 
controversy in the literature \cite{BREPLY} on whether the Mathieu 
equation gives a reliable approximation in the studies of parametric
resonance during inflaton decay. 

\section*{Acknowledgments}

This work was partly supported by the Research Committee of the University of Athens and the Hellenic State Scholarships Foundation (IKY).

\end{document}